\documentclass{aip-cp}

\usepackage[numbers]{natbib}
\usepackage{rotating}
\usepackage{graphicx}
\usepackage[T1]{fontenc}

\begin{document}

\title{Chiral Symmetry Breaking out of QCD Effective Locality}

\author[aff1]{Thierry Grandou\corref{cor1}}
\author[aff2]{Ralf Hofmann}
\eaddress{hofmann@thphys.uni-heidelberg.de}
\author[aff3]{Peter H. Tsang}
\eaddress{peter\_tsang@brown.edu}

\affil[aff1]{Universit\'e de Nice-Sophia Antipolis, Institut Non Lin\'eaire de Nice, UMR CNRS 7335; 1361 routes des Lucioles, 06560 Valbonne, France}
\affil[aff2]{Institut f\"ur Theoretische Physik,
Universit\"at Heidelberg,
Philosophenweg 16,
69120 HEIDELBERG}
\affil[aff3]{Physics Department, Brown University, Providence, RI 02912, USA}
\corresp[cor1]{Corresponding author: Thierry.Grandou@inphyni.cnrs.fr} 

\maketitle

\begin{abstract}
The $QCD$ non-perturbative property of \textit{Effective Locality} whose essential meaning has been disclosed recently, is here questioned about the chiral symmetry breaking phenomenon, one of the two major issues of the non-perturbative phase of $QCD$. As a first attempt, \textit{quenching} and the \textit{eikonal approximation} are used so as to simplify calculations which are quite involved. Chiral symmetry breaking appears to be realised in close connection to the Effective Locality mass scale, $\mu^2$, as could be expected.
\end{abstract}

\section{INTRODUCTION}
Contrary to its perturbative domain, where $QCD$ was indeed discovered and formulated, the non-perturbative sector of $QCD$ remains somehow \textit{terra incognita} where Lattice approaches are taken to provide the only reliable information. Remarkable \textit{analytic} attempts do exist though such as the MIT \textit{Bag Model} \cite{bag}, the \textit{Stochastic Vacuum Model} \cite{Hans}, the \textit{Light Cone} approach within the \textit{AdS/QCD} holographic correspondence \cite{Brodsky}, while a series of non-relativistic and relativistic quark models can be quoted also.
 
More recently a non-perturbative, exact and new property of the $QCD$ fermionic Green's functions has been discovered and given the name \textit{Effective Locality} \cite{QCD6}. It can be phrased as the following:
\par\medskip
\textit{For any fermionic $2n$-point Green's functions and related amplitudes, the
full gauge-fixed sum of cubic and quartic gluonic interactions,
fermionic loops included, results in a local contact-type interaction. This local interaction is
mediated by a tensorial field which is antisymmetric both in Lorentz and
color indices. Moreover, the resulting sum appears to be fully gauge-fixing independent, that is, gauge-invariant.}
\par\medskip
Some applications have been worked out of this highly non-trivial property and right from \textit{tree-level}, they have been found to comply with the expected features of the non-perturbative sector of $QCD$ so far as we can know it. A(n almost) linear \textit{dynamical} quark confinement potential has been derived, an interesting estimation of a `model pion' mass, the \textit{Deuteron} (Jastrow's) potential, and thanks to a non-perturbative redefinition of renormalisation, a successful description of $proton-proton$ scattering at ISR and now LHC energies has been achieved \cite{QCD5'}.
\par\medskip
That is, qualitative `tree-level' consequences of effective locality appear promising. These derivations, that began with easier and quicker cases, relying on \textit{quenching} and \textit{eikonal} aproximations, can (and must) be improved. Likewise they should also be extended to other examples in order to explore the relevance of this new property to the non-perturbative domain of QCD.

\par\medskip

At the same time though, aside from its formal derivation, the meaning of this new non-perturbative $QCD$ property remained largely obscure and it is only until recently that a profound aspect of Effective Locality could be brought to light, as the non-perturbative realisation of non-abelian gauge invariance \cite{mpla}. This characterisation of Effective Locality is obtained out of the full non-approximate QCD theory and shows, in particular, that the long issue of \emph{Gribov's copies}, closely related to the perturbative structure of QCD does not apply to its non-perturbative sector. Definitely not a trivial consequence of Effective Locality.
\par\medskip
Now, the two major issues of non-perturbative $QCD$ that must be faced are \textit{confinement} and \textit{chiral symmetry breaking}, regardless of how these two non-perturbative phenomena may be intertwined. If Effective Locality is a real and sound property of the non-perturbative sector of the $QCD$ theory, it must allow one to derive chiral symmetry breaking, it cannot be otherwise.


\section{EFFECTIVE LOCALITY: A REMINDER}

Without loss of generality, things can be illustrated on the basis of a 4-point fermionic Green's function. Then, two fermionic propagators $G_F(x_1,y_1|A)$ and $G_F(x_2,y_2|A)$ of the quark fields within a background gauge field $A_\mu$ are to be represented with the help of a \emph{Fradkin representation} such as (mixed case),
\begin{eqnarray}\label{Gfrad}
{\langle p|{G}_{F}[A] |y \rangle} &=&  e^{-i p \cdot y} \  i
\int_{0}^{\infty}\mathrm{d}s \, e^{-is m^2} \ e^{- \frac{1}{2} \mathrm{Tr}\,{\ln{\left( 2h \right)}} }\nonumber\\ \nonumber && \times \int{\mathrm{d}[u]} {\left\{ m - i \gamma \cdot \left[p - g A(y-u(s)) \right] \right\}} \ e^{\frac{i}{4} \int_{0}^{s}{\mathrm{d}s' \, [u'(s')]^{2} } } \ e^{i p \cdot u(s)}\nonumber\\  && \times  \left( e^{g \int_{0}^{s}{\mathrm{d}s' \sigma \cdot \mathbf{F}(y-u(s'))}} \ e^{-ig \int_{0}^{s}{\mathrm{d}s' \, u'(s') \cdot \mathbf{A}(y-u(s'))}} \right)_{+}\,,
\end{eqnarray}with,
\begin{equation} h(s_1,s_2)=s_1\Theta(s_2-s_1)+s_2\Theta(s_1-s_2)\ ,\ \ \ h^{-1}(s_1,s_2)=\frac{\partial}{\partial s_{1}} \frac{\partial}{\partial s_{2}} \delta(s_1-s_2).\end{equation}
In Equation (\ref{Gfrad}), the 4-vector $u(s)$ is the Fradkin variable, while in the last line the $+$-subscript indicates an $s'$-Schwinger-proper-time ordering of the expression between parenthesis. Clearly, Fradkin's representations cannot be thought of as being of a very simple usage. But they are exact. Recently, they have even been used to re-derive some standard results of Quantum Mechanics in a certainly involved but successful way \cite{mpla}.
\par
A representation like (\ref{Gfrad}) produces a cumbersome structure of an exponential of an exponential. This is why it is necessary to bring Equation (\ref{Gfrad}) down by means of functional differentiations with respect to the Grassmannian sources $\bar{\eta},\eta$ and deal with $2n$-point fermionic Green's functions. Accordingly, in contrast to the field strength formulation dual to the pure Yang Mills case \cite{Reinhardt}, the property of Effective Locality is referred to the behavior of fermionic Green's functions rather than that of the generating functional itself. Of course, this is no restriction of generality.
\par
In Equation (\ref{Gfrad}) though, the linear and quadratic $A^a_\mu$-field dependences are contained within a time-ordered exponential which prevents functional differentiations to operate in a simple way. This can be circumvented at the expense of introducing two extra field variables so as remove the $A^a_\mu$-field variables from the ordered exponential, for example, 
\begin{equation}\label{out}
\left(e^{ig\,p^\mu\!\int_{-\infty}^{+\infty} {\rm{d}}s\,A^a_\mu(y-sp)\,T^a}\right)_+=\mathcal{N_+}\int {\rm{d}}[\alpha]\int{\rm{d}}[\Omega]\, e^{-i\!\int_{-\infty}^{+\infty} {\rm{d}}s\,\,\Omega^a(s)[\alpha^a(s)-gp^\mu A^a_\mu(y-sp)]}\left(e^{i\int_{-\infty}^{+\infty}{\rm{d}}s\,\alpha^a(s)T^a}\right)_+
\end{equation}where $\mathcal{N_+}$ is a normalisation constant. Equation (\ref{out}) defines an \emph{eikonal} approximation of Equation (\ref{Gfrad}) which is used here to offer a simple derivation of the property of interest where the approximation of quenching is applied (ignoring thus the Fradkin representation of the functional $L(A)$). 
\par
Now, not until the full integrations on the two extra field variables $\alpha^a$ and $\Omega^a$ are brought to completion, can one be assured to deal with the proper Fradkin representation of $G_F[A]$. At this level of approximations though and in the strong coupling limit, $g\gg 1$, it is a fortunate circumstance that these extra dependences can be integrated out in an exact way thanks to the \emph{Random Matrix} calculus \cite{QCD6}.
\par
Omitting for short the details of integrations on Schwinger proper times, Fradkin's variables and the four extra fields $\alpha^a_i$ and $\Omega^a_i$, $i=1,2$, one obtains a result of the following form
\begin{eqnarray}\label{ELEQ}
\prod_{i=1}^{2}\int\mathrm{d}s_i \int\mathrm{d}u_i(s_i)\int\mathrm{d}\alpha_i(s_i)\int\mathrm{d}\Omega_i(s_i)\,\left(\ddots\right) \int{\mathrm{d}[\chi] \, e^{ \frac{i}{4} \int{ \chi^{2} }} } \, \left. e^{\mathfrak{D}_{A}^{(0)}} \, e^{{+ \frac{i}{2} \int{ A ^a_\mu\, K^{\mu\nu}_{ab}\, A^b_\nu} }} \, e^{i\int{Q^a_\mu A^\mu_a } }\right|_{\ A \rightarrow 0 }\,,
\end{eqnarray}
where $K^{\mu\nu}_{ab}$ and $Q^a_\mu$ represent the quadratic and linear $A^a_\mu$-field dependences, respectively,
\begin{equation}\label{QK}
K_{\mu\nu}^{ab}=gf^{abc}\chi_{\mu\nu}^c+\left({{D}_{\mathrm{F}}^{(0)}}^{-1}\right)_{\mu \nu}^{a b}\,, \ \ \ Q^a_\mu =-\partial^\nu\chi^a_{\mu\nu}+g[R^a_{1,\mu}+R^a_{2,\mu}]\,, \ \  \ f^{abc}\chi^c_{\mu\nu}=
(f\cdot \chi)^{ab}_{\mu\nu}\,,\end{equation}and where the $R^a_{i,\mu}$ stand for the part of Equation (\ref{out}) linear in the potential field $A^a_\mu$ \cite{Halpern},
 \begin{equation}\label{currents}
 R^a_{i,\mu}(z) = p_{i,\mu}\, \int\mathrm{d}s_i\,\Omega^a_{i}(s_i)\,\delta^4(z - y_{i} + s_i\, p_{i})\,, \ \ \ \ \ i=1,2\,. 
 \end{equation} In Equation (\ref{ELEQ}) and (\ref{QK}) the $\chi^a_{\mu\nu}$-fields are introduced so as to linearise the original non-belian $F^a_{\mu\nu}F_a^{\mu\nu}$ lagrangian term \cite{Halpern}. Note that in (\ref{currents}) the eikonal approximation has substituted $s_ip_i$ for the 
 original Fradkin field variable $u_i(s_i)$ ({\textit{i.e.}} a straight line approximation, connecting the points $z$ and $y_i$).
The $A^a_\mu$-functional operations followed by the prescription of cancelling the sources $j^a_\mu$ is now trivial and yields,
\begin{eqnarray}
& & \left. e^{-\frac{i}{2} \int{\frac{\delta}{\delta A} \cdot {D}_{\mathrm{F}}^{(0)} \cdot  \frac{\delta}{\delta A} }} \cdot e^{+ \frac{i}{2} \int{A \cdot {K} \cdot A} + i \int{A \cdot {Q} }}  \right|_{A \rightarrow 0}  = e^{-\frac{1}{2} \mathrm{Tr}\,{\ln{\left( 1- {D}_{\mathrm{F}}^{(0)} \cdot {K} \right)}}} \cdot e^{\frac{i}{2} \int{{Q} \cdot \left[ {D}_{\mathrm{F}}^{(0)} \cdot \left( 1 - {K} \cdot {D}_{\mathrm{F}}^{(0)}\right)^{\!-\!1} \right]\cdot {Q}}}.
\end{eqnarray}

\noindent On the right hand side, the kernel of the quadratic term in ${Q}_{\mu}^{a}$ is
\begin{eqnarray}\label{magics}
{D}_{\mathrm{F}}^{(0)} \cdot \left( 1 - {K} \cdot {D}_{\mathrm{F}}^{(0)}\right)^{\!-\!1} &=& {D}_{\mathrm{F}}^{(0)} \cdot \left( 1 - \left[ g f \cdot \chi + {{D}_{\mathrm{F}}^{(0)}}^{\!-\!1} \right] \cdot {D}_{\mathrm{F}}^{(0)}\right)^{\!-\!1} = - \left( g f \cdot \chi \right)^{\!-\!1},
\end{eqnarray}so that eventually,
\begin{eqnarray}\label{EL}
& & \left. e^{-\frac{i}{2} \int{\frac{\delta}{\delta A} \cdot {D}_{\mathrm{F}}^{(0)} \cdot  \frac{\delta}{\delta A} }} \cdot e^{+ \frac{i}{2} \int{A \cdot {K} \cdot A} + i \int{A \cdot {Q} }}  \right|_{A \rightarrow 0} = e^{-\frac{1}{2} \mathrm{Tr}\,{\ln{\bigl[-g{D}_{\mathrm{F}}^{(0)}\bigr]}}} \cdot \frac{1}{\sqrt{\det(f\cdot\chi)}}\cdot e^{-\frac{i}{2} \int\mathrm{d}^4z\ {{Q}(z) \cdot (gf\cdot\chi(z))^{-1}\cdot {Q}(z)}}\,,
\end{eqnarray}where the first term is a constant (possibly infinite) which can be absorbed into a redefinition of the overall normalization constant $\mathcal{N}$. 
\par
The manifestation of Effective Locality is in the last term of (\ref{EL}). While the $\left({\mathbf{D}_{\mathrm{F}}^{(0)}}^{-1}\right)_{\mu \nu}^{a b}$-piece of $K^{\mu\nu}_{ab}$ in (\ref{QK}) is \emph{non-local} it disappears from the final result so as to leave the \emph{local} structure $[gf\cdot\chi]^{-1}$ as the mediator of 
interactions between quarks, $\langle x | (gf\cdot\chi)^{-1} | y \rangle = (gf\cdot\chi)^{-1}(x) \, \delta^{(4)}(x-y)
$. This offers a simple way to look at the Effective Locality phenomenon whose detailed derivation can be seen to rely in an essential way on the non-abelian character of the gauge group. Contrary to expectations in effect, \cite{O}, this phenomenon cannot show up in the abelian case of QED. Going back to (\ref{ELEQ}), one finds a result whose final form reads as,
\begin{eqnarray}\label{ELEQ'}
\mathcal{N}\,\prod_{i=1}^{2}\int\mathrm{d}s_i \int\mathrm{d}u_i(s_i)\int\mathrm{d}\alpha_i(s_i)\int\mathrm{d}\Omega_i(s_i)\,\left(\ddots\right) \int{\mathrm{d}[\chi] \, e^{ \frac{i}{4} \int{ \chi^{2} }} } \cdot\frac{1}{\sqrt{\det(f\cdot\chi)}}\cdot e^{-\frac{i}{2} \int\mathrm{d}^4z\ {{Q}(z) \cdot (gf\cdot\chi(z))^{-1}\cdot {Q}(z)}}\,,\end{eqnarray}
and it is the integration on the $\chi^a_{\mu\nu}$-fields which, thanks to another remarkable consequence of Effective Locality, lends itself to a(n analytically continued) Random Matrix calculation \cite{QCD6}.
\par
The above structure generalises easily to $2n$-point fermionic Green's functions, see \cite{RefI} (Appendix D). Note that in the pure (euclidean) YM case, a form of effective locality was observed in an attempt at a dualization of the YM theory \cite{Reinhardt}.

\section{CHIRAL SYMMETYRY BREAKING}
What has been shown recently without approximation is that Effective Locality represents the non-perturbative realisation of non-abelian gauge invariance  \cite{mpla}. As a first attempt, the possibility that chiral symmetry breaking emerges out of Effective Locality will be examined within the eikonal and quenching approximations. One has,
  \begin{equation}\label{6}
<\bar{\Psi}(x)\Psi (y)>\,= \mathcal{N}  \int d[\chi] \, e^{\frac{i}{4}\int \chi^2} \, \left. e^{\mathfrak{D}_A^{(o)}} \, e^{ + \frac{i}{2}\int{\chi\cdot {F}} +\frac{i}{2}\int{A \cdot \left({D}_{F}^{(0)}\right)^{-1}\!\! \cdot\, A }} \, {G}_{\mathrm{F}}(x_{}, y_{}|A) \right|_{A\rightarrow 0}\end{equation}
Using Equation (\ref{Gfrad}), two contributions are obtained, with $e^{- \frac{1}{2} \mathrm{Tr}{\ln{\left( 2h \right)}} } = \mathcal{N}(s)$,

 \begin{eqnarray}\label{I}
\nonumber  im\int_{0}^{\infty}{ds \ e^{-is m^{2}}} \, \mathcal{N}(s) \nonumber \int{d[u]} \, e^{ \frac{i}{4} \int_{0}^{s}{ds' \, [u'(s')]^{2} } } \, \delta^{(4)}(x - y + u(s)) \\  \mathrm{Tr}\,\mathcal{N_+} \int\! {\rm{d}}[\alpha]\!\int\!{\rm{d}}[\Omega]\, e^{-i\!\int {\rm{d}}s'\,\,\Omega^a(s')\alpha^a(s')} \,(e^{i\int_{-\infty}^{+\infty}{\rm{d}}s\,\alpha^a(s)\lambda^a})_+
 \int d[\chi] \, e^{\frac{i}{4}\int \chi^2} \,  \left. e^{\mathfrak{D}_{A}^{(0)}} \, e^{{+ \frac{i}{2} \int{ A ^a_\mu\, K^{\mu\nu}_{ab}\, A^b_\nu} }} \, e^{i\int{Q^a_\mu A^\mu_a } }\right|_{\ A \rightarrow 0 }\,,\end{eqnarray}
and,
 \begin{eqnarray}\label{II}
\nonumber  i\int_{0}^{\infty}{ds \ e^{-is m^{2}}} \, \mathcal{N}(s)  \nonumber \int{d[u]} \, e^{ \frac{i}{4} \int_{0}^{s}{ds' \, [u'(s')]^{2} } } \, \delta^{(4)}(x - y + u(s)) \\  \mathrm{Tr}\,\mathcal{N_+} \int\! {\rm{d}}[\alpha]\!\int\!{\rm{d}}[\Omega]\, e^{-i\!\int {\rm{d}}s'\,\,\Omega^a(s')\alpha^a(s')} \,(e^{i\int_{-\infty}^{+\infty}{\rm{d}}s\,\alpha^a(s)\lambda^a})_+
 \int d[\chi] \, e^{\frac{i}{4}\int \chi^2} \,  \left. e^{\mathfrak{D}_{A}^{(0)}} \, \left[ig\gamma^\mu A_\mu^a(x)\lambda^a\right]
e^{{+ \frac{i}{2} \int{ A ^a_\mu\, K^{\mu\nu}_{ab}\, A^b_\nu} }} \, e^{i\int{Q^a_\mu A^\mu_a } }\right|_{\ A \rightarrow 0 }\,.\end{eqnarray}
Lengthy calculation are able to prove that in the chiral limit, \textit{i.e.}, $m\rightarrow 0$, the contribution of (\ref{I}) vanishes in agreement with a long known and quite intuitive result of J. Schwinger \cite{Schwinger}, which, though established in QED, states that scalar electrons/positrons do not lead to any chiral symmetry breaking effect. In Equation (\ref{II}) on the contrary the spinorial structure is not absent if not totally taken into account, and it is this contribution which is responsible for a breaking of the chiral symmetry. Carrying out the differential operation related to $e^{\mathfrak{D}_{A}^{(0)}}$ one obtains,
 \begin{eqnarray}\label{II2}
\nonumber   \int_{0}^{\infty}{\mathrm{d}s \ e^{-is m^{2}}} \, \mathcal{N}(s)\int{d[u]} \, e^{ \frac{i}{4} \int_{0}^{s}{ds' \, [u'(s')]^{2} } } \, \delta^{(4)}(x - y + u(s))\, \mathrm{Tr}\,\mathcal{N_+}\! \int\! {\rm{d}}[\alpha]\!\int\!{\rm{d}}[\Omega]\, e^{-i\!\int {\rm{d}}s'\,\,\Omega^a(s')\alpha^a(s')} \,(e^{i\int_{-\infty}^{+\infty}{\rm{d}}s\,\alpha^a(s)\lambda^a})_+\\ \times\int d[\chi] \, e^{\frac{i}{4}\int \chi^2} \, \frac{1}{\sqrt{\det(f\cdot\chi)}}e^{-\frac{i}{2} g\int\mathrm{d}^4z\ {{R}(z) \cdot (f\cdot\chi(z))^{-1}\cdot {R}(z)}}[g\gamma^\mu\lambda^a][(f\cdot\chi(x))^{-1}R(x)]^a_\mu \end{eqnarray}with $R^a_{\mu}(z) =p_\mu \int_0^s\mathrm{d}s'\,\Omega^a_{}(s')\,\delta^4(z - y + u(s'))$ replacing $Q^a_\mu$ in the strong coupling limit. A theorem in Wiener functional space may be invoked so as to deal with the expression,
\begin{equation}
\int_{0}^{\infty}\!\!{\rm{d}s}\int_{0}^{s}\!\!{\rm{d}s'}\,\delta^{(4)}(x - y + u(s))\,\delta^{(4)}(x - y + u(s'))=\int_{0}^{\infty}{\rm{d}s}\int_{0}^{s}{\rm{d}s'}\,\delta^{(4)}(x - y + u(s))\,\delta^{(4)}(u(s')-u(s))
\end{equation} replacing it with the result,
\begin{equation}
{1\over |\mathrm{u}'_{3}(0)|  |{\mathrm{u}}'_{0}(0)|}\,\frac{\mu^2}{\pi}\,\delta^{(4)}(x - y + u(0))
\end{equation}where $\mu^2$ is the mass attached to Effective Locality. In order to proceed with a long and involved calculation it is convenient to pass to a $N=D\times (N_c^2-1)$-dimensional space \cite{Halpern} and to redefine variables accordingly. For example, $\gamma^\mu\otimes \lambda^a=N^{a+n\mu}=N^i$ with $n=N_c^2-1$, $\mu$ the ordinary Lorentz index $\mu=0,1,2,3$ and $1\leq i\leq 32$. In the same way one defines the $N$-vectors $V^i=p^\mu\otimes \Omega^a$, ${\hat{\alpha}}=(1,1,1,1)\otimes {\mathbf{\alpha}}$ and ${\hat{\lambda}}=(1,1,1,1)\otimes {\mathbf{\lambda}}$ where $\lambda$ is the set of Gellmann matrices. In this construction the correspondance $i\leftrightarrow (\mu, a)$ is one-to-one \textit{modulo} $n$. Taking advantage of the \textit{adjoint representation} relation $f^{abc}=i(T^a)^{bc}$ the term $f\cdot\chi$ is transformed into a real symmetric traceless $N\times N$-matrix $M=-if\cdot\chi$ with spectrum $S\!pM=\{\xi_1,\dots,\xi_N\}$. Another important calculational step to be taken is the one of the analytic continuation of the \textit{Random Matrix} formalism \cite{QCD6}. 
This allows one to transform the second line of Equation (\ref{II2}) into a form,
\begin{equation}
 \int [{\rm{d}}\mathcal{O}]\  N^i \frac{1}{i}\frac{\delta}{\delta\, (\mathcal{O}{V})^i} \int_{-\infty}^{+\infty}\prod_{i=1}^{N/2}\ \frac{{\rm{d}}\xi_i}{{\sqrt{-\xi_i^2}}} \, \prod_{i=1}^{N/2}2|\xi_i|\, \prod_{1\leq i<j}^{N/2} (\xi^2_i-\xi^2_j)^2  e^{-{i\over 8N_c}\, \xi_i^2}\ \,{e^{\frac{i}{2} g\frac{\mu^2{\sqrt{\Delta}}}{\pi Ep}
\ { \frac{{V}^i{V}^i-{V}^{N-i+1}{V}^{N-i+1}}{\xi_i}}}}
\end{equation}
where $[\mathrm{d}\mathcal{O}]$ summarizes the Haar integration measure over the orthogonal group $O_N(I\!R)$ \cite{RefI}. Lengthy calculations then show that each monomial of the \textit{Vandermonde} determinant term $\prod_{1\leq i<j}^{N/2} (\xi^2_i-\xi^2_j)^2$ can be integrated exactly, the result expressed in terms of \textit{Meijer} special functions, and yields a non-zero contribution to the order parameter $<\bar{\Psi}(x)\Psi (x)>$. Results come out proportional to $\mu^3$, the effective mass scale, while decreasing as the quark field momentum increases as it should, but in a way which differs from the perturbative \textit{Asymptotic Freedom}. As the overall sum of monomials is non zero, chiral symmetry breaking seems to be realised at this order of approximations. Derivations and details of these dependences will be displayed as well as attempts to estimate the sum of all monomial contributions, relying possibly on the Wigner semi-circle approximation or the analyticity properties of the Meijer functions.
\section{CONCLUSION}
Effective Locality is an exact property of the full generating functional of $QCD$ and besides encouraging phenomenological applications, an important meaning of this property has been recently disclosed \cite{mpla}. The Effective Locality property of $QCD$ is inherent to the non-perturbative sector of this theory and this is why it should allow one to give some account of the two major issues of this terra incognita, \textit{confinement} and \textit{chiral symmetry breaking}. Calculations performed at quenching and eikonal approximations seem to support the latter effect which comes out closely related to the Effective Locality mass scale, as could be somewhat expected.



\end{document}